\documentclass[aps,pra,reprint,superscriptaddress]{revtex4-1}

\usepackage{subfigure}
\usepackage{float}

\usepackage{amsmath}
\usepackage{graphicx}
\usepackage{color}

\begin{document}
\title{Experimental entanglement-assisted quantum delayed-choice experiment}
\date{\today}
\author{Tao Xin}
\author{Hang Li}
\author{Bi-Xue Wang}
\affiliation{State Key Laboratory of Low-dimensional Quantum Physics and Department of Physics, Tsinghua University, Beijing 100084, China}
\affiliation{The Innovative Center of Quantum Matter, Beijing 100084, China}
\affiliation{Tsinghua National Laboratory of Information Science and Technology,  Beijing 100084, China}

\author{Gui-Lu Long}
\email[Correspondence and request materials should be addressed to G.L.L.: ]{gllong@tsinghua.edu.cn}
\affiliation{State Key Laboratory of Low-dimensional Quantum Physics and Department of Physics, Tsinghua University, Beijing 100084, China}
\affiliation{The Innovative Center of Quantum Matter, Beijing 100084, China}
\affiliation{Tsinghua National Laboratory of Information Science and Technology,  Beijing 100084, China}

\begin{abstract}
The puzzling properties of quantum mechanics, wave-particle duality, entanglement and superposition, were dissected experimentally at past decades. However, hidden-variable (HV) models, based on three classical assumptions of   wave-particle objectivity, determinism and independence, strive to explain or even defeat them. The development of quantum technologies enabled us to test experimentally the predictions of quantum mechanics and HV theories. Here, we report an experimental demonstration of an entanglement-assisted quantum delayed-choice scheme using a liquid nuclear magnetic resonance quantum information processor. This scheme we realized is based on the recently proposed scheme [Nat. Comms. 5:4997(2014)], which gave different results for quantum mechanics and HV theories. In our experiments,  the intensities and the visibilities of the interference are in consistent the theoretical prediction of quantum mechanics. The results imply that a contradiction is appearing when all three assumptions of HV models are combined, though any two of the above assumptions are compatible with it.
\end{abstract}
\maketitle

\section{Introduction}
The properties of quantum mechanics is proverbially obscure \cite{wheeler} because the predictions of quantum theory are very different from our classical expectations. Meanwhile, HV models can reproduce the quantum mechanics results, making it another alternative explanation,  and even attempting to challenge quantum theory \cite{bell,peres}. The emergence of quantum technologies \cite{chuang} makes it possible to distinguish quantum mechanics and HV theories. For example, HV theories lead to conflict with the predictions of quantum theory when all three classical assumptions are combined in the 'entanglement-assisted quantum delayed-choice experiment' \cite{wave}, which originates from the investigation of wave-particle duality.

Wave-particle duality is one of most puzzling properties of quantum theory, which can be nicely illustrated by the classical Wheeler delayed-choice (WDC) experiment \cite{greenberger}. In this setup (Fig. 1(a)), it includes the first beam splitter BS1 splitting the  trajectory of each incoming photon into two optional paths, and the lower and upper paths are respectively named  as $|0\rangle$ and $|1\rangle$ states. A phase shifter in the upper path introduces a variable phase shift $\varphi$. The second beam splitter BS2 is controlled by a random number generator (RNG) that decides the insertion or removal of BS2. Detectors D0 and D1 are used to measure the photon. An interference pattern depending on the phase shift $\varphi$ is observed in a closed interferometer with BS2 inserted, which implies that the photon goes through both paths simultaneously and interfers with itself at BS2.  Hence the photon is like a wave. In an open interferometer without BS2, there is no interference pattern, which means that the photon traveled only through one path,  and the photon is like particle.

Recently, Jacques \textit{et al} \cite{Jacques} have demonstrated this proposal. Their result of delay-choice experiment shows that the behavior of the photon in the interferometer depends on the removal or insertion of the beam splitter BS2.

A modified version of Wheeler's experiment has been proposed by Ionicioin and Torno \cite{terno}, which can observe the simultaneous presence of particle and wave behaviors. In this setup (Fig. 1(b)), BS2 is replaced by a quantum beam splitter (QBS) which is controlled by the superposition state of a two-level system. The delayed-choice experiment with quantum control is named quantum delayed-choice experimental, which has been recently implemented in several different systems \cite{Kaiser,NMR,accaise,tang2012realization,Peruzzo}. In another development, wave-particle duality can be clearly comprehended via the Realistic Interpretation (REIN) of quantum mechanics proposed by one of us, which has been demonstrated in the corresponding Encounter-Delayed-Choice (EDC) experiment \cite{long2014}.

Based on the quantum versions of WDC, entanglement-assisted quantum delayed-choice(EAQDC) experiment was proposed \cite{wave}, which not only demonstrates wave-particle duality, but also provides the possibility of testing the predictions of quantum mechanics and HV theories. This is due to the fact that the intuitive ideas of determinism \cite{brand}, wave-particle objectivity \cite{terno} and locality \cite{pirono} are mutually inconsistent when the HV description is extended to this set-up \cite{celeri}.

\begin{figure*}[!htp]
\centering
\subfigure[]{
\label{Fig.sub.1}
\includegraphics[width=0.32\textwidth]{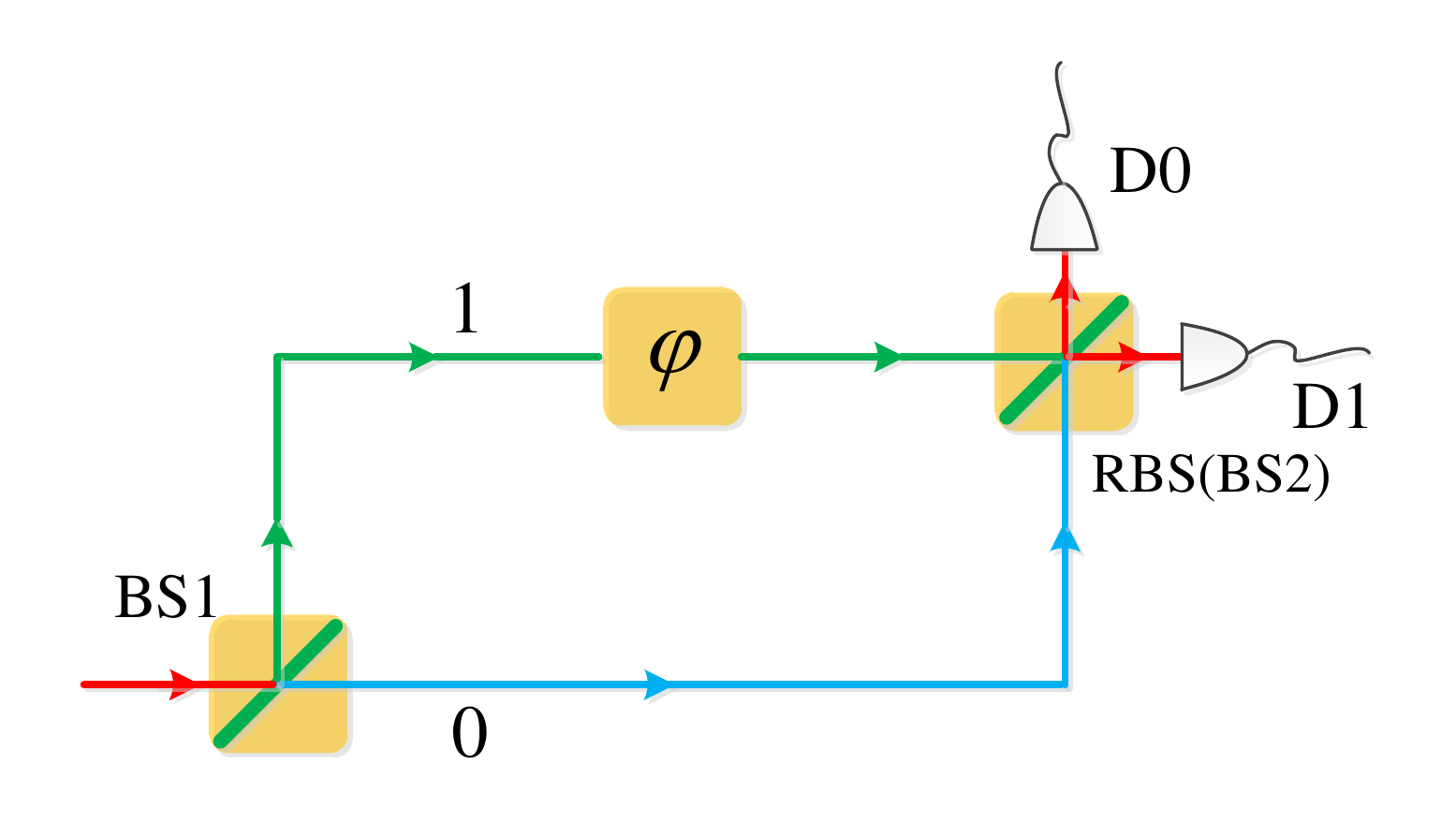}}
\subfigure[]{
\label{Fig.sub.3}
\includegraphics[width=0.32\textwidth]{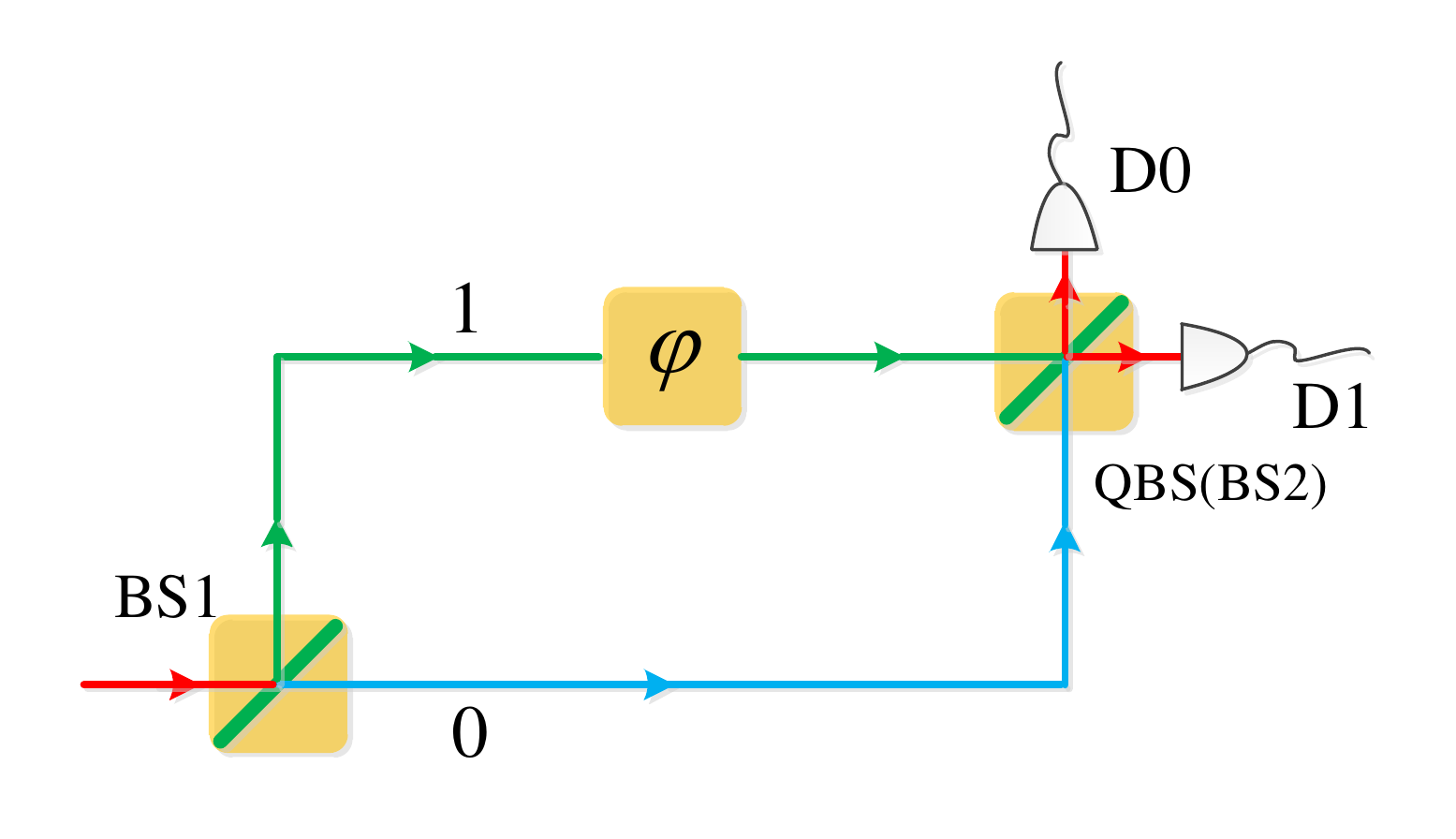}}
\subfigure[]{
\label{Fig.sub.4}
\includegraphics[width=0.32\textwidth]{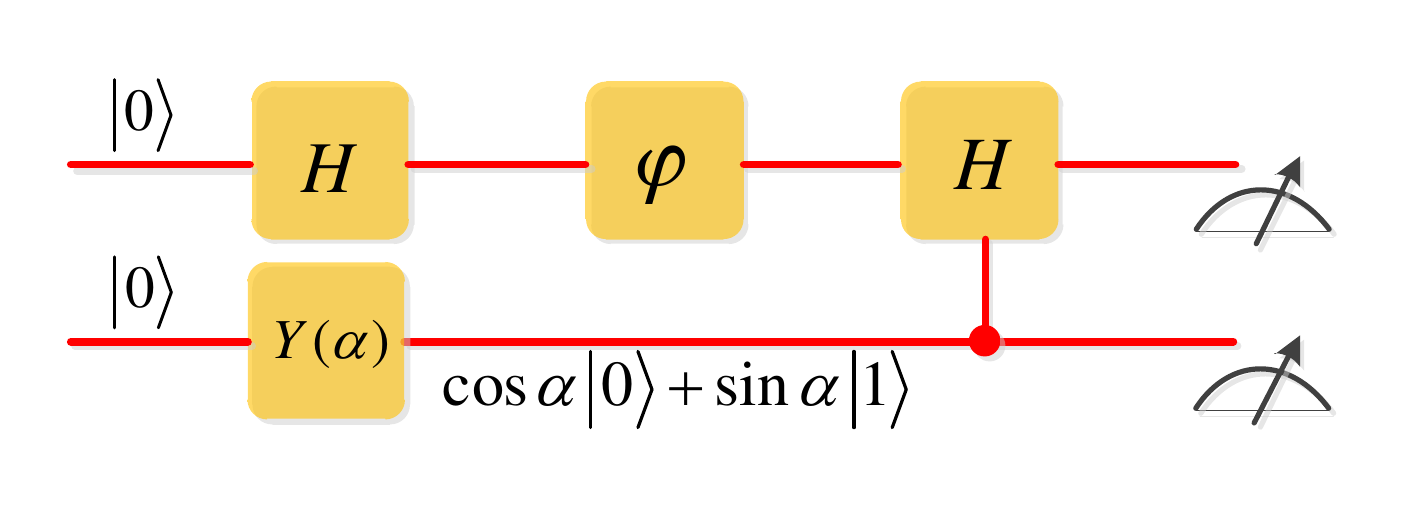}}
\subfigure[]{
\label{Fig.sub.5}
\includegraphics[width=0.41\textwidth]{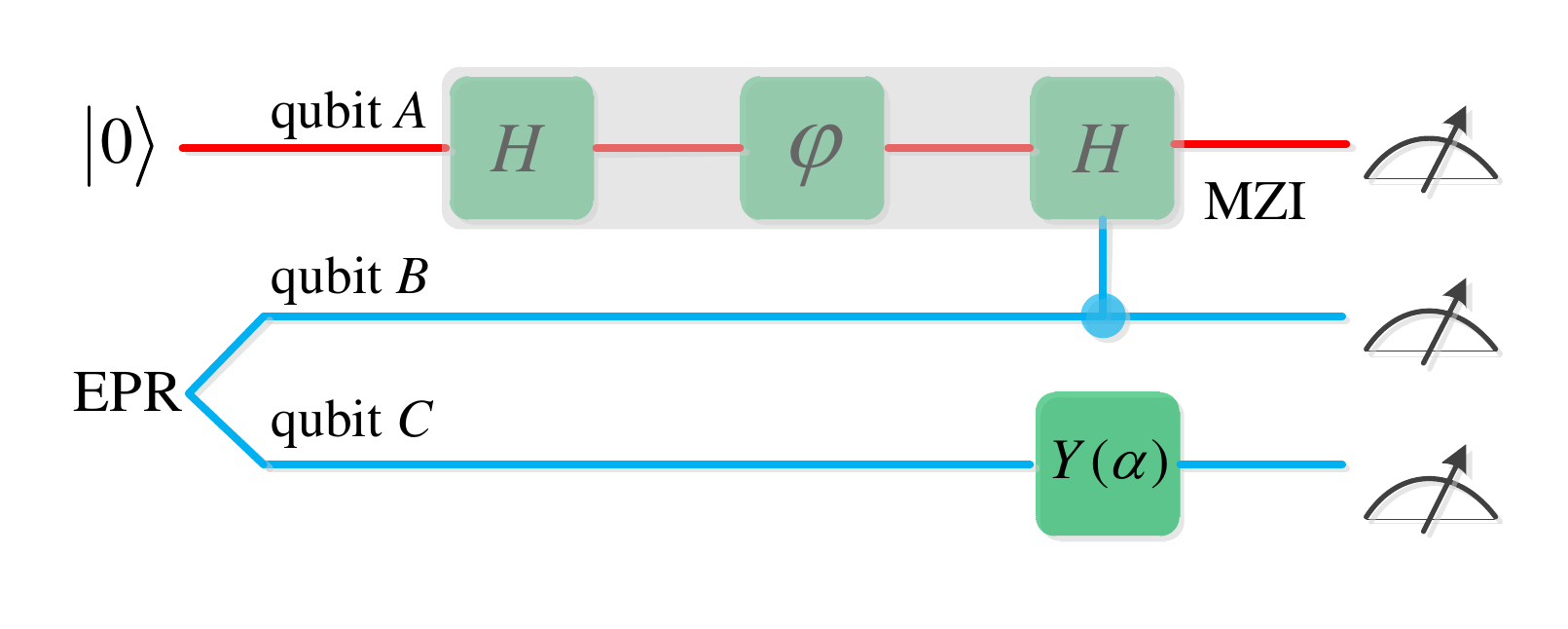}}
\subfigure[]{
\label{Fig.sub.6}
\includegraphics[width=0.4\textwidth]{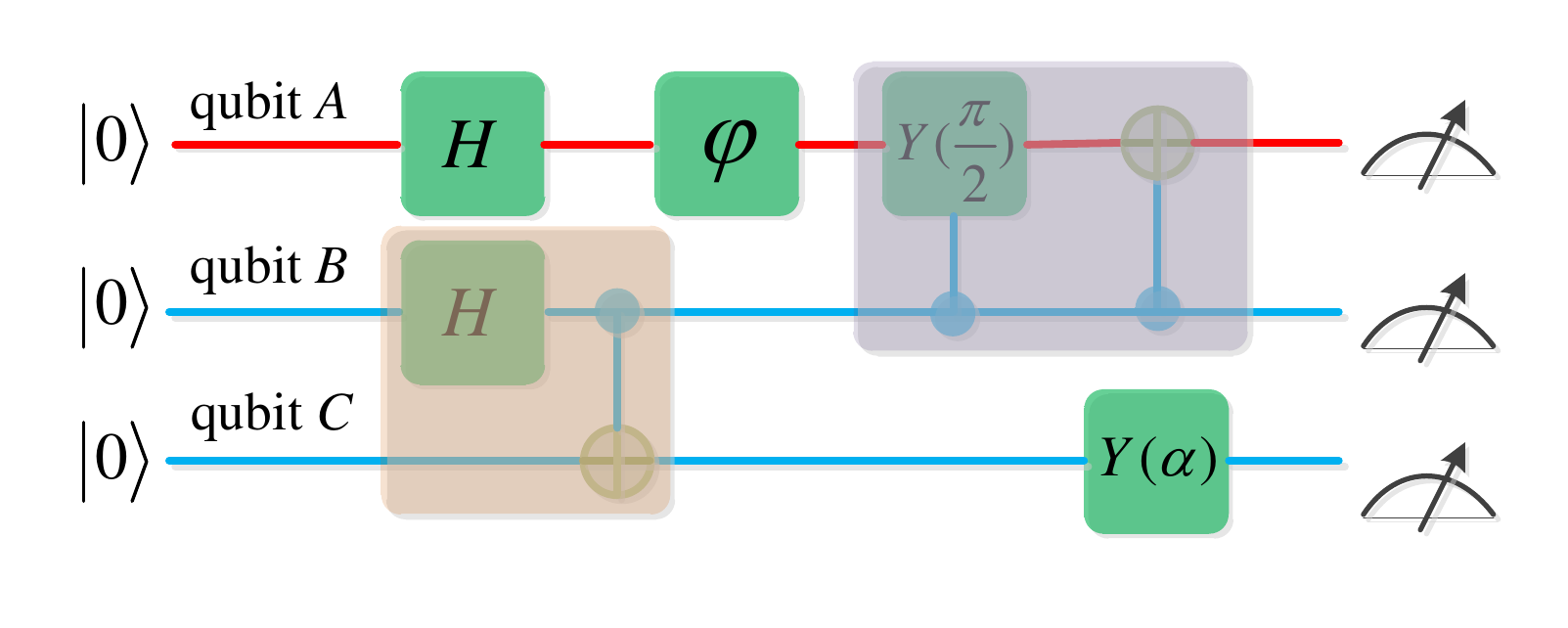}}
\caption{The Wheeler Delayed-Choice experiment. (a) The classical WDC setup. If BS2 is present, it is a 'closed interferometer'. If BS2 is absent, it is an 'open interferometer'. (b) The Wheeler delayed-choice with quantum control. QBS is a beam splitter which is controlled by a quantum system in superposition. (c) The equivalent quantum circuit of Fig. \ref{fig1}(b). First qubit is initially prepared in  $|0\rangle$. Second qubit is an auxiliary quantum system initially prepared in the superposition state $\cos{\alpha}|0\rangle+\sin\alpha|1\rangle$. $Y(\alpha)=\exp\left(i\alpha {\sigma_y}\right)$ is used to prepare the state of the ancillary qubit initially prepared in the $|0\rangle$ state. (d) The entanglement-assisted quantum delayed-choice experiment. EPR is an Einstein-Podolsky-Rosen pair. (e) Feasible quantum circuit of entanglement-assisted quantum delayed-choice experiment. Here, $H$ is the Hadamard gate and $\varphi$ is a phase shifter.}
\label{fig1}
\end{figure*}

As shown in Fig. 1(d), it is a three-qubit system. $A$ is a work qubit which is initially prepared in the $|0\rangle$ state. $B$ and $C$ are two ancillary qubits which are initially prepared the $\left(\sqrt{\eta}|00\rangle+\sqrt{1-\eta}|11\rangle\right)$ state, for $\eta=\dfrac{1}{2}$, $BC$ is a maximally entangled EPR pair. In our experiments, $\eta=\dfrac{1}{2}$. $A$ enters a MZI in which second beamsplitter is quantum-controlled by qubit $B$. $C$ undergoes a rotation around $Y$ axis  $Y(\alpha)=\exp\left(i\alpha {\sigma_y}\right)$ followed by a measurement in the computational basis. The final density matrix $\rho=|\psi\rangle\langle\psi|$,
\begin{equation}
\begin{aligned}
|\psi\rangle={\dfrac{1}{\sqrt{2}}(\cos\alpha|p\rangle|0\rangle+\sin\alpha|w\rangle|1\rangle)_{AB}|0\rangle_{C}-}\\
{\dfrac{1}{\sqrt{2}}(\sin\alpha|p\rangle|0\rangle-\cos\alpha|w\rangle|1\rangle)_{AB}|1\rangle_{C}.}
\end{aligned}
\end{equation}
in which $|p\rangle=\dfrac{1}{\sqrt{2}}(|0\rangle+e^{i\varphi}|1\rangle), |w\rangle=e^{i\varphi/2}(\cos\dfrac{\varphi}{2}|0\rangle-i\sin\dfrac{\varphi}{2}|1\rangle)$.
\begin{table}[H]
\centering
\begin{tabular}{c|c|c}
\hline

     &QM prediction & HV prediction \\
\hline
$I_{A|c=0}$ &$\dfrac{1}{4}\cos^2\alpha+\dfrac{1}{2}\sin^2\alpha\cos^2\dfrac{\varphi}{2}$ & $\dfrac{1}{4}+\dfrac{1}{2}\cos^2\dfrac{\varphi}{2}$ \\
\hline
$I_{A|c=1}$ &$\dfrac{1}{4}\sin^2\alpha+\dfrac{1}{2}\cos^2\alpha\cos^2\dfrac{\varphi}{2}$ & $\dfrac{1}{4}+\dfrac{1}{2}\cos^2\dfrac{\varphi}{2}$ \\
\hline
$V_{A|c=0}$&${\sin^2\alpha}$ & $\dfrac{1}{2}$ \\
\hline
$V_{A|c=1}$ &${\cos^2\alpha}$& $\dfrac{1}{2}$ \\
\hline
\end{tabular}
 \caption{The intensities and the visibilities predicted by quantum mechanics and HV theories in EAQDC experiments. $I_{A}=Tr(\rho_A|0\rangle\langle 0|)$, with $\rho_A=Tr_{BC}|\psi\rangle\langle\psi|$. If the outcome of measurement qubit $C$ is $c=0 (c=1)$, $I_{A}$ is postselected, resulting in $I_{A|c=0}(I_{A|c=1})$. $V_{A|c=0}(V_{A|c=1})$ is the corresponding visibility according to $V=(I_{max}-I_{min})/(I_{max}+I_{min})$. }
 \label{table1}
 \end{table}

The intensities and visibilities of the interference predicted by quantum mechanics (QM) and HV theories are shown in Table \ref{table1}.  Therefore, the predictions of HV theories are in contrast with those of quantum mechanics when $\cos2\alpha\neq0$. Readers can acquire details of calculation in Ref.\cite{wave}.

In this article, we present a feasible experimental scheme of entanglement-assisted quantum delayed-choice experiment and realize it via a liquid NMR quantum information processor \cite{Rev,havel}. Our results show that the intensities $(I_{A|c=0},I_{A|c=1})$ and visibilities $(V_{A|c=0},V_{A|c=1})$ are in accordance with quantum mechanics description. Therefore, it is concluded that determinism, local independence and wave-particle objectivity are not always compatible with quantum mechanics. In other words, HV theories fail to explain results of entanglement-assisted quantum delayed-choice experiment.

\section{Results}
\subsection{Circuit design of entanglement-assisted WDC}
A three-qubit quantum system of NMR sample is often initially prepared in the $|000\rangle$ state. As shown in Fig. 2(a), we design a feasible quantum circuit so that qubits $BC$ initially prepared in the $|00\rangle$ state is transformed into an EPR pair (the light-orange in the lower-left corner of Fig. 1(e)). qubits $BC$ undergoes the Hadamard gate on qubit $B$ followed by the C-NOT gate in which $B$ is a control qubit and $C$ is a target qubit, finally transformed into the $\dfrac{1}{\sqrt{2}}(|00\rangle+|11\rangle)$ state. Fig.2(b) illustrates that the control-Hadamard gate is decomposed into a combination of $C-R_y(\pi/2)$ and $C-R_x(\pi)$,   $U_{C-H}=U_{C-R_x(\pi)}U_{C-R_y(\pi/2)}$. The final feasible quantum circuit of entanglement-assisted quantum delayed-choice experiment is shown in Fig. 1(e).
\begin{figure}[H]
\centering
\subfigure[]{
\label{Fig.sub.7}
\includegraphics[width=0.25\textwidth]{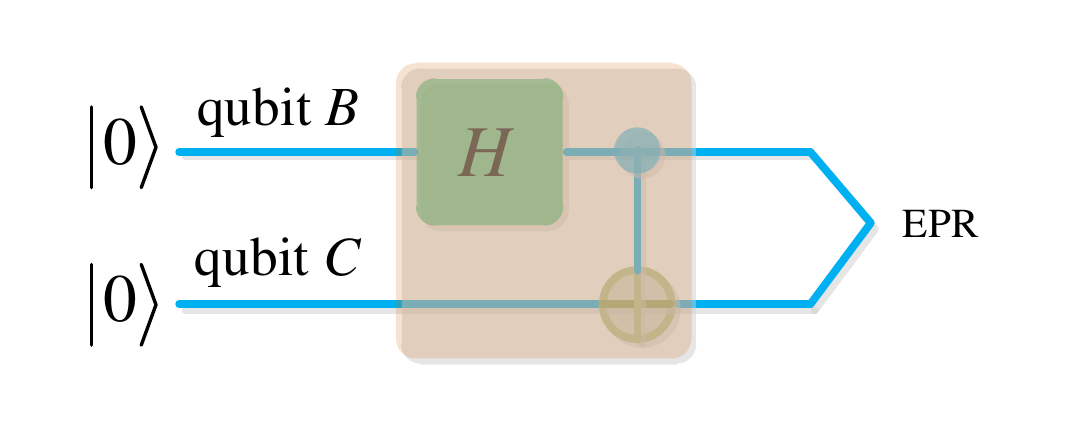}}
\subfigure[]{
\label{Fig.sub.8}
\includegraphics[width=0.3\textwidth]{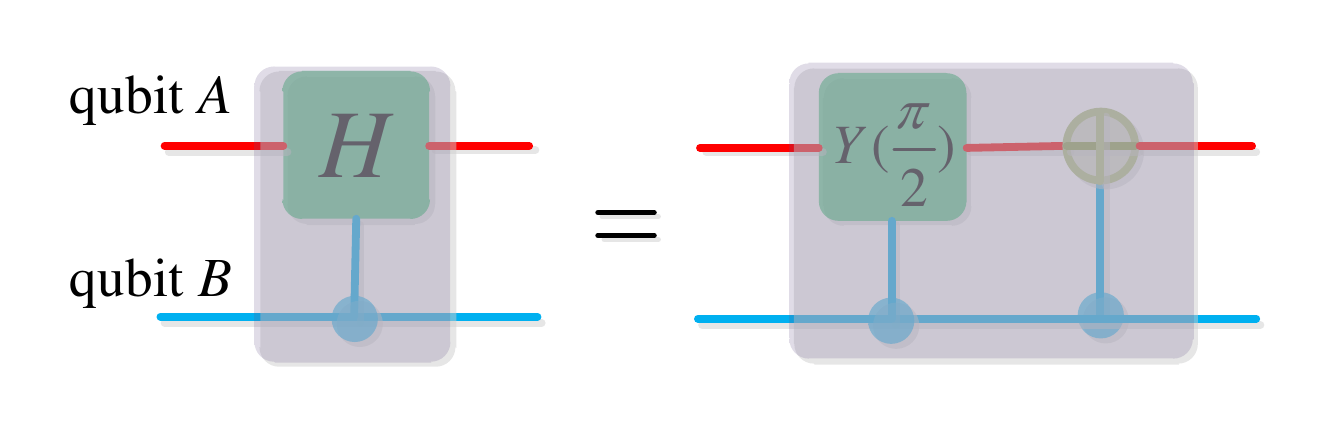}}
\caption{Parts of quantum circuit design. (a)Quantum circuit of generating an EPR pair. (b) The equivalent circuit of the control-Hadamard gate. There is no operation on qubit $A$ if qubit $B$ is in the $|0\rangle$ state. Qubit $A$ successively undergoes a rotation $R_y(\pi/2)=e^{-i\pi\sigma_y/4}$ and a rotation $R_x(\pi)=e^{-i\pi\sigma_x/2}$ if qubit $B$ is in the $|1\rangle$ state.}
\label{Fig.2}
\end{figure}

\subsection{The pulse sequence design and experimental implementation}
The NMR pulse sequence of entanglement-assisted quantum delayed-choice experiment is illustrated by Fig. 3(a). This design requires a three-qubit quantum system, a work qubit ($^{13}$C) and two ancillary qubits ($^1$H and $^{19}$F).
\begin{figure*}[!htp]
\centering
\subfigure[]{
\label{Fig.sub.71}
\includegraphics[width=0.97\textwidth]{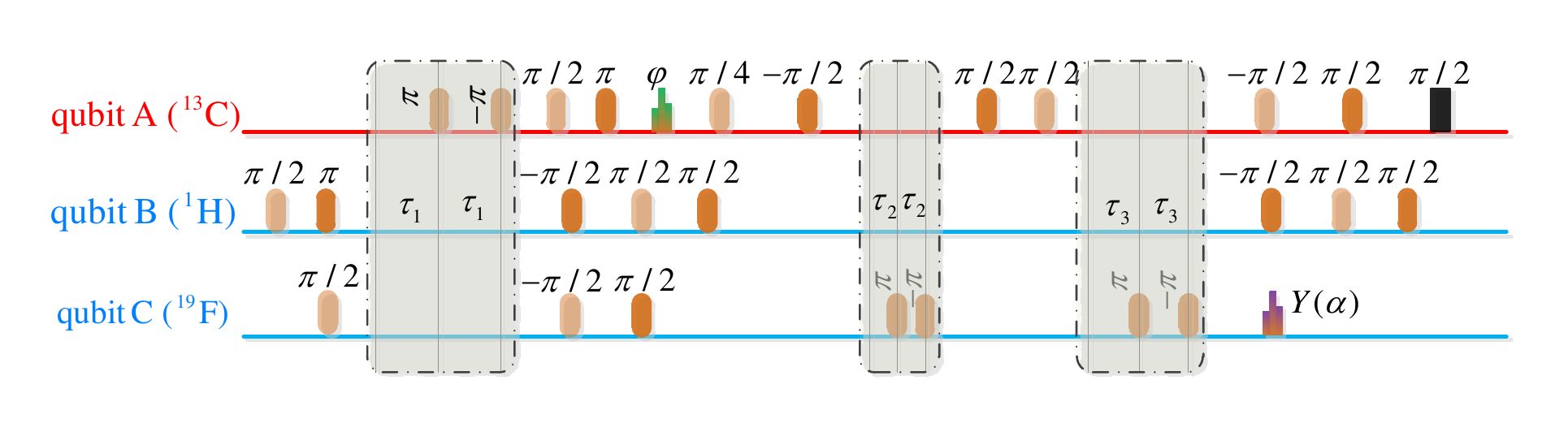}}
\subfigure[]{
\label{Fig.sub.81}
\includegraphics[width=0.92\textwidth]{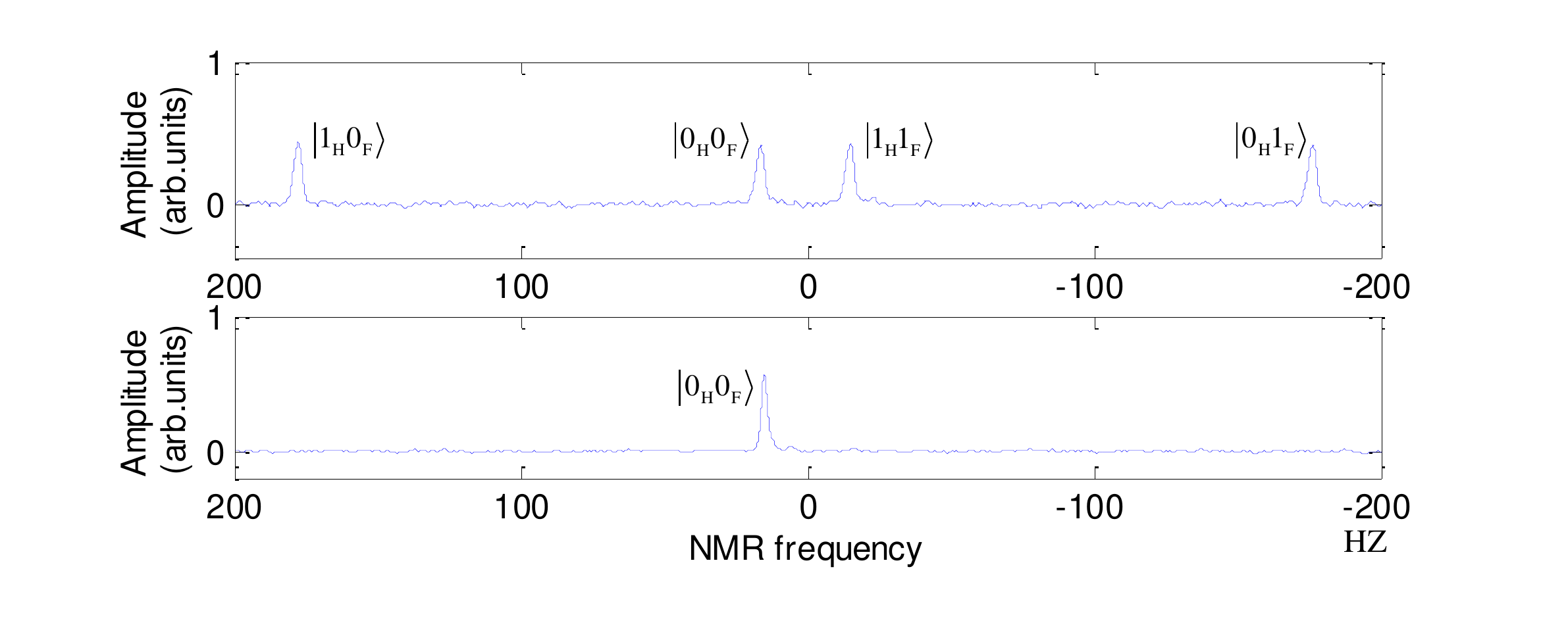}}
\caption{The entire sequence and experimental spectra of $^{13}$C. (a)The entire pulse sequence corresponds to entanglement-assisted quantum delayed-choice experiment. Light-orange and deep-orange pulses are hard pulses. Deep-orange pulses represent X-pulse: $R_x(\theta)=e^{-i\theta\sigma_x/2}$. Light-orange pulses represent Y-pulse: $R_y(\theta)=e^{-i\theta\sigma_y/2}$. $\theta$ is a rotation angle on the top of each pulse. Ladder-pulses are shaped pulses which are strongly modulated pulses \cite{fortunato,mahesh}. First and second ladder-pulse respectively act as a phase shifter $\varphi$ on qubit $A$ and $Y(\alpha)=\exp\left(i\alpha {\sigma_y}\right)$ on qubit $C$. $J_{ij}$ is the $J$-coupling constant between the $i$th and $j$th nucleus. In our experiments, $J_{ab}$=160.8HZ, $J_{bc}$=47.6HZ. The $J$-coupling evolution time $\tau_{1}=\dfrac{1}{4J_{bc}}$, $\tau_{2}=\dfrac{1}{8J_{ab}}$, $\tau_{3}=\dfrac{1}{4J_{ab}}$. Each dotted rectangle is for realizing two-qubit gates. First, second, and third dotted rectangle respectively realize the evolution $e^{-i\pi\sigma_z^b\sigma_z^c/4}$, $e^{-i\pi\sigma_z^a\sigma_z^b/8}$, and $e^{-i\pi\sigma_z^a\sigma_z^b/4}$. $\dfrac{\pi}{2}$ detection pulse is shown in black pulse. (b)Experimental spectra of $^{13}$C. States of other two qubits decide the position of the peaks. }
\label{Fig.3}
\end{figure*}
All the gates on the work qubit and the ancillary qubits were decomposed into the combination of the hard pulses. The phase-shift $\varphi$ operator on the qubit $A$ and $Y(\alpha)$ operator on qubit $C$ were realized via shaped pulses which were achieved by calculating Hibert-Schmidt fidelity between the desired operator and the experimental operator for different radio frequency (rf) amplitude and phase \cite{suter,khan}. Finally, an average fidelity of over 0.995 was achieved. Other one-qubit gates were realized via a combination of one-qubit rotation pulses $R_x(\theta)=e^{-i\theta\sigma_x/2}$ and $R_y(\theta)=e^{-i\theta\sigma_y/2}$. Two-qubit gates were decomposed into one-qubit rotations and the evolutions under the spin-spin coupling.

In our experiments, the sample consisted of $^{13}$C-unlabeled diethyl-fluoromalonate dissolved in $d6$-acetone is used as three-qubit NMR quantum computer processor. $^{13}$C, $^1$H, and $^{19}$F respectively acts as the work qubit and the two ancillary qubits. These spin-$\dfrac{1}{2}$ nuclei are placed in the strong magnetic field resulting in the Zeenman splitting. Spin-up $|\uparrow\rangle$ (Spin-down $|\downarrow\rangle$ ) is defined as $|0\rangle$ state ($|1\rangle$ state). Energy level will be splitted again due to two-body interaction in the NMR sample. Transmitting the resonance pulses leads to energy level transition, which is detected by the probe of NMR spectrometer. For instance, as shown in the upper part of Fig.3(b), the NMR spectrum of $^{13}$C is obtained through a rotation $R_y(\pi/2)=e^{-i\pi\sigma_y/4}$ on the equilibrium state of diethyl-fluoromalonate. The peaks are labeled by corresponding states of the other two qubits ($^1$H and $^{19}$F) in the molecule. This is the theoretical foundation of using the NMR sample as the three-qubit quantum system that we need. Molecular structure and properties of the NMR sample are listed in Methods.

The equilibrium state of NMR system does not correspond to a pseudo-pure state. Therefore, we redistribute the populations using the spatial averaging technique \cite{coryar} to prepare the desired pseudo-pure state \cite{ensemble,pure,chuang1998}:
\begin{equation}
\rho_0\propto(\dfrac{I}{2}+I_z^1)\otimes(\dfrac{I}{2}+I_z^2)\otimes(\dfrac{I}{2}+I_z^3)-\dfrac{1}{8}I\otimes I\otimes I
\end{equation}
where $I_z^k=\dfrac{1}{2}\sigma_z^k$ ($k$ is the $k$th nuclear) and $I$ is a $2\times 2$ unitary matrix. The identity part of $\rho_0$ does not give rise to NMR signals and is ignored. Experimental spectrum of $^{13}$C after the preparation of pseudo-pure state is illustrated in the lower part of Fig. 3(b). Only one peak in this spectrum implies that NMR sample is prepared in the $|000\rangle$ state. Details of the preparation of the pseudo-pure state are in Methods.

\subsection{Experimental conclusions}
Our purpose is to acquire a probability distribution $I_{A|c=0}$ ($I_{A|c=1}$) when qubits $^{13}$C and $^{19}$F are in the $|00\rangle$ ($|01\rangle$) state. Final density matrix $\rho$ provides some desired information about $I_{A|c=0}$ and $I_{A|c=1}$. Our calculation shows that $I_{A|c=0}$ is the sum of the first and the third diagonal element of $\rho$, $I_{A|c=0}=\rho(1,1)+\rho(3,3)$. $I_{A|c=1}$ is the sum of the second and the firth diagonal element of $\rho$, $I_{A|c=1}=\rho(2,2)+\rho(4,4)$. Therefore, the complete density matrix tomography is not necessary. The most general density matrix of a three-qubit system is of the form:
\begin{equation}
\rho=\sum_{i,i,k=0}^3c_{ijk}\sigma_i \otimes \sigma_j \otimes \sigma_k.
\end{equation}
with the unknown constant $c_{ijk}$. The diagonal part of $\rho$ is
\begin{equation}
\begin{aligned}
\rho_{diag}={c_0III+c_1\sigma_zII+c_2I\sigma_zI+c_3II\sigma_z+}\\
{c_4\sigma_z\sigma_zI+c_5I\sigma_z\sigma_z+c_6\sigma_zI\sigma_z+c_7\sigma_z\sigma_z\sigma_z.}
\end{aligned}
\end{equation}
where $c_{0}=\dfrac{1}{8}$ and $c_{i}$ ($i=1,2,...,6,7$) is the unknown constant.

Our calculation shows that $I_{A|c=0}=2(c_0+c_1+c_3+c_6)$ and $I_{A|c=1}=2(c_0+c_1-c_3-c_6)$. For the given phase shift $\varphi$ and rotation $Y(\alpha)$, implementing two experiments is enough to acquire the constants $c_1$, $c_3$, and $c_6$. Firstly, Experimental spectrum of $^{13}$C obtained by directly observing $^{13}$C after implementing the entire pulses sequence shown in Fig. 3(a) gives four signals proportional to $c_1-c_4+c_6-c_7$, $c_1+c_4+c_6+c_7$, $c_1-c_4-c_6+c_7$ and $c_1+c_4-c_6-c_7$. The constant $c_3$ can be obtained by directly observing $^{19}$F, but it is not allowed in our NMR spectrometer. Fortunately, $c_3$ is acquired by doing a SWAP gate to transfer the state of $^{19}$F to the work qubit ($^{13}$C) and then observing $^{13}$C, instead of observing $^{19}$F directly. The SWAP gate is realized using the shaped pulse whose total time is 7.5$ms$. The spectrum similar with the lower part of Fig. 3(b) is obtained by implementing a SWAP operator after preparing the pseudo-pure state, implying that this operator is accurate. Analogously, the spectrum of $^{13}$C after doing the SWAP operator behind the entire sequence gives four signals proportional to $c_3-c_5+c_6-c_7$, $c_3+c_5+c_6+c_7$, $c_3-c_5-c_6+c_7$ and $c_3+c_5-c_6-c_7$. Therefore, from these four signals one can precisely determine all three unknown constants $c_1$, $c_3$, and $c_6$. and obtain the intensities $I_{A|c=0}$ and $I_{A|c=1}$.

\begin{figure*}[!htp]
\centering
\subfigure[]{
\label{Fig.sub.9}
\includegraphics[width=0.8\textwidth]{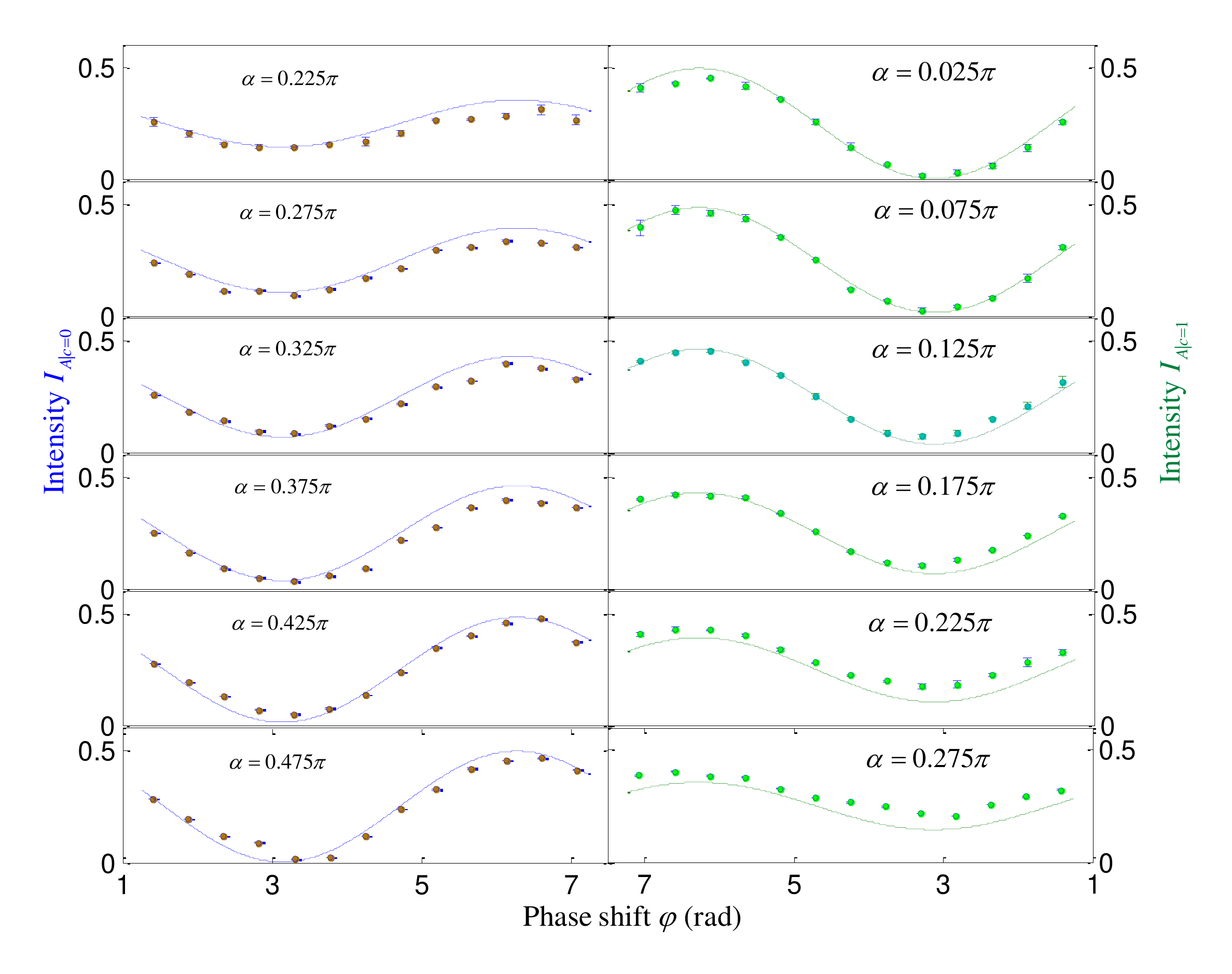}}
\subfigure[]{
\label{Fig.sub.10}
\includegraphics[width=0.8\textwidth]{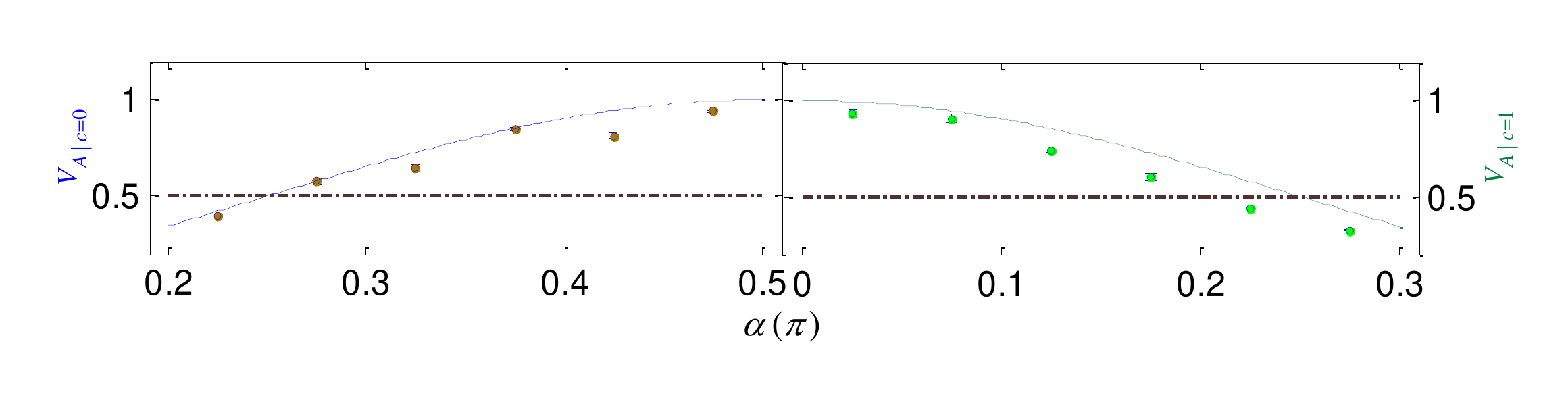}}
\caption{Experimental intensities and visibilities. (a)Intensities $I_{A|c=0}$ and $I_{A|c=1}$ versus phase $\varphi$ for different values of $\alpha$. The quantum mechanical predictions are shown by solid lines. This experiment was repeated four times. The average values of experimental results are shown by symbols. Almost of error bars are covered by the symbols because experimental values have the small standard deviation of less than 0.03. (b)The visibilities $V_{A|c=0}$ and $V_{A|c=1}$ versus rotation angle $\alpha$. QM predicts that $V_{A|c=0}$ (blue) and $V_{A|c=1}$ (cyan) respectively vary as $\sin^2\alpha$ and $\cos^2\alpha$. In HV theories, the visibilities does not distinguish $c=0,1$ cases. The dotted line $V^{HV}=0.5$ illustrates this. }
\label{Fig.4}
\end{figure*}

Experimental intensities $I_{A|c=0}$ and $I_{A|c=1}$ are shown in Fig. 4(a). The intensities were measured for six values of $\alpha$. For each given value of $\alpha$, thirteen values of the phase shift $\varphi$ were selected. The theoretical value predicted by quantum mechanics are also shown in Fig. 4(a). Some factors such as  imperfect pulse operations lead to some experimental values drifting off the theoretical curve. More detailed error analysis are presented in Discussions. Fig. 4(b) illustrates the visibilities $V_{A|c=0}$ and $V_{A|c=1}$ calculated at different values of $\alpha$.
According to figure 4, our results present a general agreement between the quantum mechanical predictions and the experiments. It indicates that HV theories fail to explain the results of entanglement-assisted quantum delayed-choice experiment.

\section{Discussions}
In this section, we first discuss the sources of the errors in the experimental implementation of the pulse sequence shown in Fig. 3(a). Excluding the state initialization, the number of radio-frequency pulses is about 30 with experiment duration being about 28$ms$ for the case that a SWAP gate is included. Total error is magnified, although each pulse has small error. Therefore, a dominant source of errors originated from the imperfect selective $\dfrac{\pi}{2}$-pulse used.  The signal loss is also caused by some experimental factors like $T_2$ effects, the measurement (determination of the constants $c_1$,$c_3$ and $c_6$) errors and the inhomogeneity of the external magnetic field. It is acceptable that some experimental results drift off the theoretical curves, but the accuracy of experimental results is enough to reveal the incompatibility between the basic assumptions of HV theories and quantum mechanics. Meanwhile, the $^{13}$C in diethyl-fluoromalonate used is un-labeled, as a result, the desired signal is vulnerable to the environmental noise. Signal-noise ratio can be improved by using $^{13}$C-labeled diethyl-fluoromalonate or the experiment being performed many more times. If these factors can be overcome, a better agreement between quantum mechanical predictions and the experimental results can be achieved.

For $^{13}$C-unlabeled diethyl-fluoromalonate, rotating the state of the $^{19}$F nuclei is more imprecise than other two qubits ($^{13}$C and $^1$H), and hence $^{19}$F can be set to the logical qubit which needs the least operations to achieve more precise manipulation. The $J$-coupling constant between $^{13}$C and $^{19}$F ($J=-192.48$Hz) is maximum in all three $J$-coupling constants. It indicates that the $J$-coupling evolution between $^{13}$C and $^{19}$F contributes to decease $T_2$ effects, because the experiment duration is proportional to $\frac{1}{2J}$, but the calculations of the intensities $I_{A|c=0}$ and $I_{A|c=1}$ will become more complicated.

We have presented an experimental demonstration of entanglement-assisted quantum delayed-choice scheme using NMR spectrometer which provides a prefect platform to study the relationship between quantum mechanics and HV theories. Our experiment not only demonstrated continuous morphing of quantum systems between wave-like and particle-like behavior, but also tested the predictions of quantum mechanics and HV theories. An entanglement assisted version of quantum delayed-choice experiment can be also carried out in other quantum systems, such as, the entangled photon system \cite{photon}.

\section{Methods}
\subsection{Experimental system and NMR sample}
Experiments are carried out on a BRUKER AVANCE III 400MHZ NMR spectrometer with diethyl-fluoromalonate dissolved in $d6$-acetone at 295.0K as the quantum information processor. $^{13}$C, $^1$H, and $^{19}$F in the diethyl-fluoromalonate molecule respectively act as the work qubit and the two ancillary qubits matching to Fig. 3(a) of the main text. The natural Hamiltonian of the three-qubit system is as follow,
\begin{equation}
H_{NMR}^3=-\sum_{i=1}^3\omega_i\sigma_{zi}+\sum_{i<j,i=1}^3\dfrac{\pi J_{ij}}{2}\sigma_{zi} \sigma_{zj} .
\end{equation}
where $\omega_i$ is the Larmor frequency of the $i$th nuclei and $J_{ij}$ is the $J$-coupling constant between the $i$th and the $j$th nuclei. Fig. 5 illustrates the molecular structure and parameters of diethyl-fluoromalonate. The Larmor frequency of  $^{13}$C,$^1$H and $^{19}$F are 101MHZ, 63MHZ and 376MHZ, respectively. Therefore, it is feasible that the hard pulse can selectively only excites the transition of one nucleus without changing the states of other qubits. In our experiments, the duration time of each pulse is
tens of microseconds, and hence the causing undesired free-evolution can be ignored.
\begin{figure}[!htp]
\centering
\includegraphics[width=0.4\textwidth]{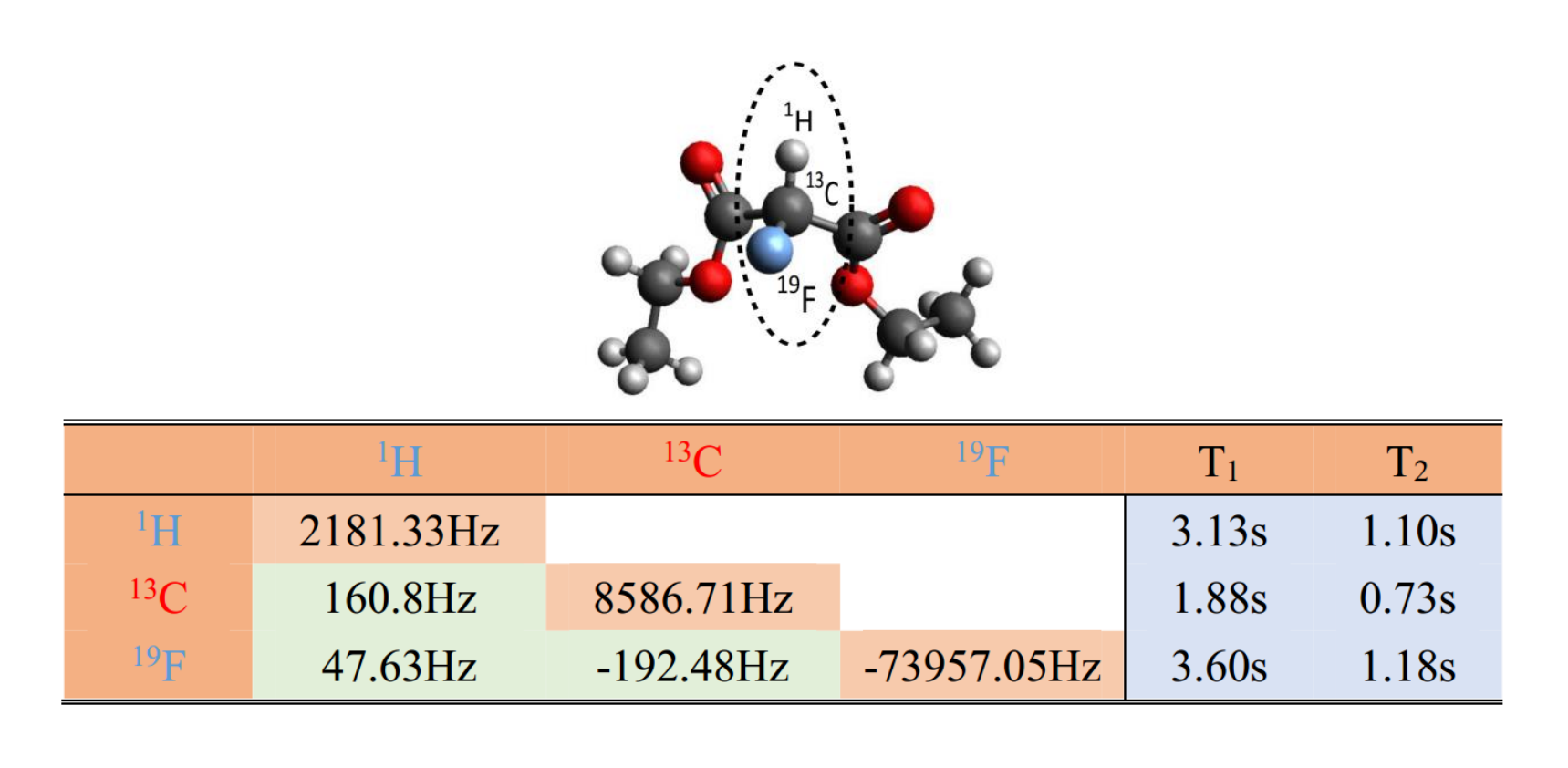}
\caption{Molecular structure and properties of NMR sample. $^{13}$C, $^1$H and $^{19}$F comprise a three-qubit NMR quantum processor. The chemical shifts and $J_{ij}$-coupling constants of this molecule are given as diagonal elements and off-diagonal elements in the table of Fig. 5. T$_1$ and T$_2$ are the longitudinal and transversal relaxation time, respectively, which can be obtained by the standard inversion recovery and Hahn echo sequences \cite{hahn}.}
\label{Fig.5}
\end{figure}

\subsection{Measurement of $\dfrac{\pi}{2}$-pulse power and state initialization}
In our experiments, almost all of unitary operators were realized using hard pulses. Whether hard pulses are prefect or imperfect directly determines the realization of the entire pulse sequence in Fig. 3(a). Experimentally, we applied four $\dfrac{\pi}{2}$-pulses on the $^{13}$C nuclei and then observing the thermal equilibrium spectrum of the $^{13}$C shown in the corresponding block of Fig. 3(b). Each $\dfrac{\pi}{2}$-pulse power was continuously assigned without changing the duration time of the pulse until four signals faded away. It implied that the desired $\dfrac{\pi}{2}$-pulse of the $^{13}$C nuclei was found. $\dfrac{\pi}{2}$-pulse power of the $^1$H ($^{19}$F) nuclei was obtained by transferring the polarization of $^1$H ($^{19}$F) to the work qubit ($^{13}$C) and observing $^{13}$C using analogous method, instead of observing $^1$H ($^{19}$F) directly. Finally, all three unknown $\dfrac{\pi}{2}$-pulse power values can be precisely determined.

The three-qubit system consisted of $^{13}$C, $^1$H and $^{19}$F from diethyl-fluoromalonate needs to be prepared in the initial state $\dfrac{1}{\sqrt{2}}|0\rangle_A(|00\rangle+|11\rangle)_{BC}$ to implement entanglement-assisted quantum delayed-choice experiment. We first create the pseudo-pure state (the equation(2)) using a spatial averaging method \cite{coryar}. The process of preparing the state is shown in Fig. 6. At last, to prepare the state$\dfrac{1}{\sqrt{2}}|0\rangle_A(|00\rangle+|11\rangle)_{BC}$, a quantum circuit (Fig.2(a)) we designed is applied to qubits $B$ and $C$.
\begin{figure}[H]
\centering
\includegraphics[width=0.29\textwidth]{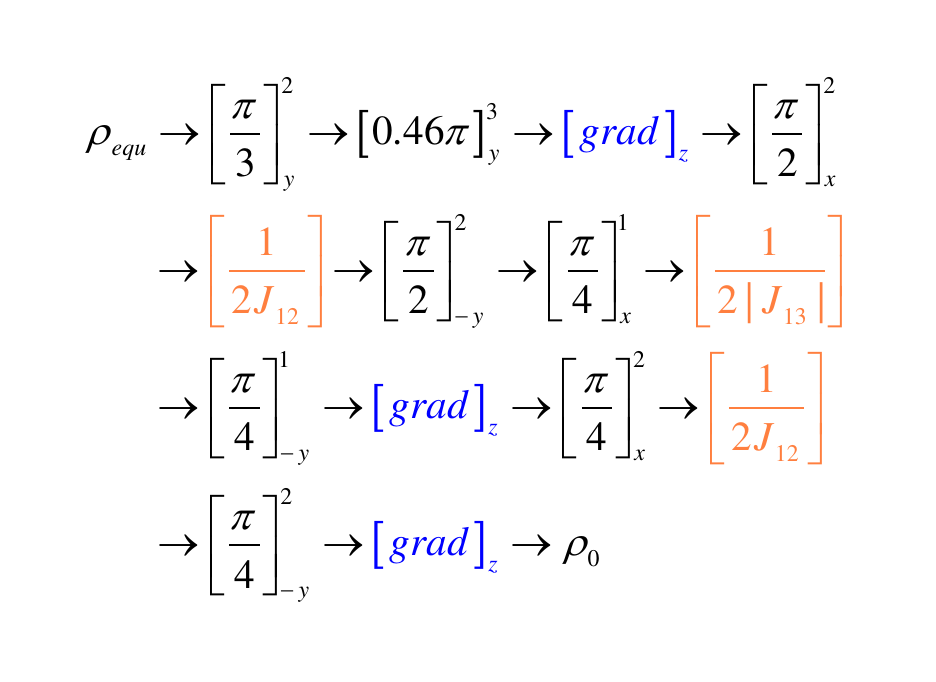}
\caption{The process of the spatial averaging method. The symbol $[\theta]_{\vec{n}}^m$ means the $m$th qubit is rotated a angle $\theta$ about a direction $\vec{n}$. The symbols $\left[ \dfrac{1}{2J_{12}}\right]$ and $\left[ \dfrac{1}{2|J_{13}|}\right]$ respectively represent the $J$-coupling evolution $e^{-i\pi\sigma_z^1\sigma_z^2/4}$ and $e^{-i\pi\sigma_z^1\sigma_z^3/4}$. The pulsed-field gradient which averages out all the coherences and retains only the diagonal part of the density matrix is shown by the symbols $[grad]_z$. $\rho_{equ}$ is the thermal equilibrium and $\rho_0$ is the desired pseudo-pure state. }
\label{Fig.6}
\end{figure}

\subsection{The determination of the constants $c_1$,$c_3$ and $c_6$}
The final density matrix of the three-qubit system $\rho$ was acquired after accomplishing entanglement-assisted quantum delayed-choice experiment, which can be reconstructed completely via Quantum State Tomography (QST) \cite{lee2002quantum}. In QST theory, the density matrix can be estimated from ensemble averages of a set of observables. For the one-qubit system, the observable set is $\lbrace\sigma_i\rbrace$ ($i=0,1,2,3$). For the three-qubit system, the observable set is $\lbrace\sigma_i\otimes\sigma_j\otimes\sigma_k\rbrace$ ($i,j,k=0,1,2,3$). Here $\sigma_0=I$, $\sigma_1=X$, $\sigma_2=Y$, $\sigma_3=Z$. In our experiments, consider that $I_{A|c=0}=2(c_0+c_1+c_3+c_6)$ and $I_{A|c=1}=2(c_0+c_1-c_3-c_6)$, in which $c_0$, $c_1$, $c_3$ and $c_6$ are the coefficients of the equation(4), and hence complete density matrix tomography is not necessary. All we need is to preform two experiments. Firstly, the unitary transforms $e^{-i\pi Y/4}$, $I$ and $I$ are respectively implemented on qubit $A$ ($^{13}$C), qubit $B$ ($^1$H) and qubit $C$ ($^{19}$F), and the work qubit ($^{13}$C) needs to be observed to present four peaks. The integration of the peaks relative to one of the pseudo-pure state are denoted as $f_1$, $f_2$, $f_3$ and $f_4$ from left to right, and then follow these steps to calculate the constants $c_1$ and $c_6$:
\begin{equation}
\begin{aligned}
{c_1=\dfrac{1}{8}(f_1+f_2+f_3+f_4),}\\
{c_6=\dfrac{1}{8}(f_1+f_2-f_3-f_4).}
\end{aligned}
\end{equation}
where $\dfrac{1}{8}$ is the normalized coefficient. Secondly, in order to acquire the constant $c_3$, a SWAP gate which transfer the state of qubit $C$ ($^{19}$F) to the work qubit $A$ ($^{13}$C) is applied to the three-qubit system before implementing the unitary transform $e^{-i\pi Y/4}\otimes I\otimes I$. Analogously, four peaks are given (denoted as $f_1'$, $f_2'$, $f_3'$ and $f_4'$, respectively). Then the constants $c_3$ and $c_6$ can be determined by calculating the following equations:
\begin{equation}
\begin{aligned}
{c_3=\dfrac{1}{8}(f_1'+f_2'+f_3'+f_4'),}\\
{c_6=\dfrac{1}{8}(f_1'+f_2'-f_3'-f_4').}
\end{aligned}
\end{equation}

Therefore, the two experiments above all determine the constant $c_6$, but we give preference to the former. The SWAP gate in the second experiment was realized using the shaped pulse. The number of the step of the pulse is $1500$ and the duration of each step is 5$\mu s$. So the total duration of the shaped pulse is 7.5$ms$. The latter is more affected by some factors like de-coherence and environmental noise than the former. From the equations (6) and (7), all three unknown constants $c_1$,$c_3$ and $c_6$ can be precisely determined.

\vskip 12pt

\newpage

\textbf{Acknowledgements}

This work is supported by the National Natural Science
Foundation of China under Grants Grants No. 11175094, 11474181 and No. 91221205, the National Basic Research Program of China under Grant No.
2011CB9216002.  We appreciate IQC, University of Waterloo, for providing the software package for NMR pulse optimization and simulation.


\begin{thebibliography}{10}
\expandafter\ifx\csname url\endcsname\relax
  \def\url#1{\texttt{#1}}\fi
\expandafter\ifx\csname urlprefix\endcsname\relax\def\urlprefix{URL }\fi
\providecommand{\bibinfo}[2]{#2}
\providecommand{\eprint}[2][]{\url{#2}}

\bibitem{wheeler}
\bibinfo{author}{Wheeler, J.~A.} \& \bibinfo{author}{Zurek, W.}
\newblock \emph{\bibinfo{title}{Quantum Theory and Measurement}}
  (\bibinfo{publisher}{Princeton, NJ: Princeton University Press},
  \bibinfo{year}{1983}).

\bibitem{bell}
\bibinfo{author}{BELL, J.~S.}
\newblock \bibinfo{title}{On the problem of hidden variables in quantum
  mechanics}.
\newblock \emph{\bibinfo{journal}{Rev. Mod. Phys.}}
  \textbf{\bibinfo{volume}{38}}, \bibinfo{pages}{447--452}
  (\bibinfo{year}{1966}).

\bibitem{peres}
\bibinfo{author}{Peres, A.}
\newblock \emph{\bibinfo{title}{Quantum theory: concepts and methods}},
  vol.~\bibinfo{volume}{57} (\bibinfo{publisher}{Springer},
  \bibinfo{year}{1995}).

\bibitem{chuang}
\bibinfo{author}{Nielsen, M.~A.} \& \bibinfo{author}{Chuang, I.~L.}
\newblock \emph{\bibinfo{title}{Quantum computation and quantum information}}
  (\bibinfo{publisher}{Cambridge university press}, \bibinfo{year}{2010}).

\bibitem{wave}
\bibinfo{author}{Ionicioiu, R.}, \bibinfo{author}{Jennewein, T.},
  \bibinfo{author}{Mann, R.~B.} \& \bibinfo{author}{Terno, D.~R.}
\newblock \bibinfo{title}{Is wave-particle objectivity compatible with
  determinism and locality ?}
\newblock \emph{\bibinfo{journal}{Nature communications}}
  \textbf{\bibinfo{volume}{5}} (\bibinfo{year}{2014}).

\bibitem{greenberger}
\bibinfo{author}{Greenberger, D.}, \bibinfo{author}{Hentschel, K.} \&
  \bibinfo{author}{Weinert, F.}
\newblock \emph{\bibinfo{title}{Compendium of quantum physics: concepts,
  experiments, history and philosophy}} (\bibinfo{publisher}{Springer},
  \bibinfo{year}{2009}).

\bibitem{Jacques}
\bibinfo{author}{Jacques, V.} \emph{et~al.}
\newblock \bibinfo{title}{Experimental realization of wheeler's delayed-choice
  gedanken experiment}.
\newblock \emph{\bibinfo{journal}{Science}} \textbf{\bibinfo{volume}{315}},
  \bibinfo{pages}{966--968} (\bibinfo{year}{2007}).

\bibitem{terno}
\bibinfo{author}{Ionicioiu, R.} \& \bibinfo{author}{Terno, D.~R.}
\newblock \bibinfo{title}{Proposal for a quantum delayed-choice experiment}.
\newblock \emph{\bibinfo{journal}{Phys. Rev. Lett.}}
  \textbf{\bibinfo{volume}{107}}, \bibinfo{pages}{230406}
  (\bibinfo{year}{2011}).

\bibitem{Kaiser}
\bibinfo{author}{Kaiser, F.}, \bibinfo{author}{Coudreau, T.},
  \bibinfo{author}{Milman, P.}, \bibinfo{author}{Ostrowsky, D.~B.} \&
  \bibinfo{author}{Tanzilli, S.}
\newblock \bibinfo{title}{Entanglement-enabled delayed-choice experiment}.
\newblock \emph{\bibinfo{journal}{Science}} \textbf{\bibinfo{volume}{338}},
  \bibinfo{pages}{637--640} (\bibinfo{year}{2012}).

\bibitem{NMR}
\bibinfo{author}{Roy, S.~S.}, \bibinfo{author}{Shukla, A.} \&
  \bibinfo{author}{Mahesh, T.~S.}
\newblock \bibinfo{title}{Nmr implementation of a quantum delayed-choice
  experiment}.
\newblock \emph{\bibinfo{journal}{Phys. Rev. A}} \textbf{\bibinfo{volume}{85}},
  \bibinfo{pages}{022109} (\bibinfo{year}{2012}).

\bibitem{accaise}
\bibinfo{author}{Auccaise, R.} \emph{et~al.}
\newblock \bibinfo{title}{Experimental analysis of the quantum complementarity
  principle}.
\newblock \emph{\bibinfo{journal}{Phys. Rev. A}} \textbf{\bibinfo{volume}{85}},
  \bibinfo{pages}{032121} (\bibinfo{year}{2012}).

\bibitem{tang2012realization}
\bibinfo{author}{Tang, J.-S.} \emph{et~al.}
\newblock \bibinfo{title}{Realization of quantum wheeler's delayed-choice
  experiment}.
\newblock \emph{\bibinfo{journal}{Nature Photonics}}
  \textbf{\bibinfo{volume}{6}}, \bibinfo{pages}{600--604}
  (\bibinfo{year}{2012}).

\bibitem{Peruzzo}
\bibinfo{author}{Peruzzo, A.}, \bibinfo{author}{Shadbolt, P.},
  \bibinfo{author}{Brunner, N.}, \bibinfo{author}{Popescu, S.} \&
  \bibinfo{author}{O’Brien, J.~L.}
\newblock \bibinfo{title}{A quantum delayed-choice experiment}.
\newblock \emph{\bibinfo{journal}{Science}} \textbf{\bibinfo{volume}{338}},
  \bibinfo{pages}{634--637} (\bibinfo{year}{2012}).

\bibitem{long2014}
\bibinfo{author}{Long, G.-L.}, \bibinfo{author}{Qin, W.},
  \bibinfo{author}{Yang, Z.} \& \bibinfo{author}{Li, J.-L.}
\newblock \bibinfo{title}{Realistic interpretation of quantum mechanics and
  encounter-delayed-choice experiment}.
\newblock \emph{\bibinfo{journal}{arXiv preprint arXiv:1410.4129}}
  (\bibinfo{year}{2014}).

\bibitem{brand}
\bibinfo{author}{Brandenburger, A.} \& \bibinfo{author}{Yanofsky, N.}
\newblock \bibinfo{title}{A classification of hidden-variable properties}.
\newblock \emph{\bibinfo{journal}{Journal of Physics A: Mathematical and
  Theoretical}} \textbf{\bibinfo{volume}{41}}, \bibinfo{pages}{425302}
  (\bibinfo{year}{2008}).

\bibitem{pirono}
\bibinfo{author}{Branciard, C.}, \bibinfo{author}{Gisin, N.} \&
  \bibinfo{author}{Pironio, S.}
\newblock \bibinfo{title}{Characterizing the nonlocal correlations created via
  entanglement swapping}.
\newblock \emph{\bibinfo{journal}{Phys. Rev. Lett.}}
  \textbf{\bibinfo{volume}{104}}, \bibinfo{pages}{170401}
  (\bibinfo{year}{2010}).

\bibitem{celeri}
\bibinfo{author}{C{\'e}leri, L.~C.} \emph{et~al.}
\newblock \bibinfo{title}{Quantum control in foundational experiments}.
\newblock \emph{\bibinfo{journal}{Foundations of Physics}}
  \textbf{\bibinfo{volume}{44}}, \bibinfo{pages}{576--587}
  (\bibinfo{year}{2014}).

\bibitem{Rev}
\bibinfo{author}{Vandersypen, L. M.~K.} \& \bibinfo{author}{Chuang, I.~L.}
\newblock \bibinfo{title}{Nmr techniques for quantum control and computation}.
\newblock \emph{\bibinfo{journal}{Rev. Mod. Phys.}}
  \textbf{\bibinfo{volume}{76}}, \bibinfo{pages}{1037--1069}
  (\bibinfo{year}{2005}).

\bibitem{havel}
\bibinfo{author}{Havel, T.} \emph{et~al.}
\newblock \bibinfo{title}{Quantum information processing by nuclear magnetic
  resonance spectroscopy}.
\newblock \emph{\bibinfo{journal}{American Journal of Physics}}
  \textbf{\bibinfo{volume}{70}}, \bibinfo{pages}{345--362}
  (\bibinfo{year}{2002}).

\bibitem{fortunato}
\bibinfo{author}{Fortunato, E.~M.} \emph{et~al.}
\newblock \bibinfo{title}{Design of strongly modulating pulses to implement
  precise effective hamiltonians for quantum information processing}.
\newblock \emph{\bibinfo{journal}{The Journal of chemical physics}}
  \textbf{\bibinfo{volume}{116}}, \bibinfo{pages}{7599--7606}
  (\bibinfo{year}{2002}).

\bibitem{mahesh}
\bibinfo{author}{Mahesh, T.~S.} \& \bibinfo{author}{Suter, D.}
\newblock \bibinfo{title}{Quantum-information processing using strongly dipolar
  coupled nuclear spins}.
\newblock \emph{\bibinfo{journal}{Phys. Rev. A}} \textbf{\bibinfo{volume}{74}},
  \bibinfo{pages}{062312} (\bibinfo{year}{2006}).

\bibitem{suter}
\bibinfo{author}{Suter, D.} \& \bibinfo{author}{Mahesh, T.}
\newblock \bibinfo{title}{Spins as qubits: Quantum information processing by
  nuclear magnetic resonance}.
\newblock \emph{\bibinfo{journal}{The Journal of chemical physics}}
  \textbf{\bibinfo{volume}{128}}, \bibinfo{pages}{052206}
  (\bibinfo{year}{2008}).

\bibitem{khan}
\bibinfo{author}{Khaneja, N.}, \bibinfo{author}{Reiss, T.},
  \bibinfo{author}{Kehlet, C.}, \bibinfo{author}{Schulte-Herbr{\"u}ggen, T.} \&
  \bibinfo{author}{Glaser, S.~J.}
\newblock \bibinfo{title}{Optimal control of coupled spin dynamics: design of
  nmr pulse sequences by gradient ascent algorithms}.
\newblock \emph{\bibinfo{journal}{Journal of Magnetic Resonance}}
  \textbf{\bibinfo{volume}{172}}, \bibinfo{pages}{296--305}
  (\bibinfo{year}{2005}).

\bibitem{coryar}
\bibinfo{author}{Cory, D.~G.}, \bibinfo{author}{Price, M.~D.} \&
  \bibinfo{author}{Havel, T.~F.}
\newblock \bibinfo{title}{Nuclear magnetic resonance spectroscopy: An
  experimentally accessible paradigm for quantum computing}.
\newblock \emph{\bibinfo{journal}{Physica D: Nonlinear Phenomena}}
  \textbf{\bibinfo{volume}{120}}, \bibinfo{pages}{82--101}
  (\bibinfo{year}{1998}).

\bibitem{ensemble}
\bibinfo{author}{Cory, D.~G.}, \bibinfo{author}{Fahmy, A.~F.} \&
  \bibinfo{author}{Havel, T.~F.}
\newblock \bibinfo{title}{Ensemble quantum computing by nmr spectroscopy}.
\newblock \emph{\bibinfo{journal}{Proceedings of the National Academy of
  Sciences}} \textbf{\bibinfo{volume}{94}}, \bibinfo{pages}{1634--1639}
  (\bibinfo{year}{1997}).

\bibitem{pure}
\bibinfo{author}{Knill, E.}, \bibinfo{author}{Chuang, I.} \&
  \bibinfo{author}{Laflamme, R.}
\newblock \bibinfo{title}{Effective pure states for bulk quantum computation}.
\newblock \emph{\bibinfo{journal}{Phys. Rev. A}} \textbf{\bibinfo{volume}{57}},
  \bibinfo{pages}{3348--3363} (\bibinfo{year}{1998}).

\bibitem{chuang1998}
\bibinfo{author}{Chuang, I.~L.}, \bibinfo{author}{Gershenfeld, N.},
  \bibinfo{author}{Kubinec, M.~G.} \& \bibinfo{author}{Leung, D.~W.}
\newblock \bibinfo{title}{Bulk quantum computation with nuclear magnetic
  resonance: theory and experiment}.
\newblock \emph{\bibinfo{journal}{Proceedings of the Royal Society of London.
  Series A: Mathematical, Physical and Engineering Sciences}}
  \textbf{\bibinfo{volume}{454}}, \bibinfo{pages}{447--467}
  (\bibinfo{year}{1998}).

\bibitem{photon}
\bibinfo{author}{Kwiat, P.~G.} \emph{et~al.}
\newblock \bibinfo{title}{New high-intensity source of polarization-entangled
  photon pairs}.
\newblock \emph{\bibinfo{journal}{Phys. Rev. Lett.}}
  \textbf{\bibinfo{volume}{75}}, \bibinfo{pages}{4337--4341}
  (\bibinfo{year}{1995}).

\bibitem{hahn}
\bibinfo{author}{Hahn, E.~L.}
\newblock \bibinfo{title}{Spin echoes}.
\newblock \emph{\bibinfo{journal}{Phys. Rev.}} \textbf{\bibinfo{volume}{80}},
  \bibinfo{pages}{580--594} (\bibinfo{year}{1950}).

\bibitem{lee2002quantum}
\bibinfo{author}{Lee, J.-S.}
\newblock \bibinfo{title}{The quantum state tomography on an nmr system}.
\newblock \emph{\bibinfo{journal}{Physics Letters A}}
  \textbf{\bibinfo{volume}{305}}, \bibinfo{pages}{349--353}
  (\bibinfo{year}{2002}).

\end{thebibliography}
\end{document}